\documentclass[prd,amsmath,amssymb,showpacs,superscriptaddress,nofootinbib,twocolumn]{revtex4-2}
\usepackage{graphicx}
\usepackage{epsfig,graphics,subfigure,psfrag,amssymb}
\usepackage{lineno}
\usepackage{dcolumn}
\usepackage{bm}
\usepackage{overpic}
\usepackage{xspace}
\usepackage{rotating}
\usepackage{color}
\usepackage{multirow}
\usepackage{colortbl}
\usepackage{epstopdf}
\definecolor{blackblue}{RGB}{17,118,208}
\definecolor{deepblue}{RGB}{0,0,51}
\usepackage[colorlinks,linkcolor=blackblue,anchorcolor=blue,citecolor=blackblue,urlcolor=blackblue]{hyperref}

\newcommand{\BESIII}{BES\uppercase\expandafter{\romannumeral3}\xspace}

\begin{document}
\title{%
{\boldmath Measurement of $e^+ e^- \rightarrow \phi \eta^{\prime}$ cross sections at center-of-mass energies between 3.508 and 4.600 GeV}}

\author{\small{M.~Ablikim$^{1}$, M.~N.~Achasov$^{10,b}$, P.~Adlarson$^{68}$, M.~Albrecht$^{4}$, R.~Aliberti$^{28}$, A.~Amoroso$^{67A,67C}$, M.~R.~An$^{32}$, Q.~An$^{64,50}$, X.~H.~Bai$^{58}$, Y.~Bai$^{49}$
% ...
\\
\vspace{0.2cm}
(BESIII Collaboration)\\
\vspace{0.2cm} {\it
$^{1}$ Institute of High Energy Physics, Beijing 100049, People's Republic of China\\
% ...
$^{4}$ Bochum Ruhr-University, D-44780 Bochum, Germany\\
% ...
$^{10}$ G.I. Budker Institute of Nuclear Physics SB RAS (BINP), Novosibirsk 630090, Russia\\
% ...
$^{28}$ Johannes Gutenberg University of Mainz, Johann-Joachim-Becher-Weg 45, D-55099 Mainz, Germany\\
% ...
$^{32}$ Liaoning Normal University, Dalian 116029, People's Republic of China\\
% ...
$^{49}$ Southeast University, Nanjing 211100, People's Republic of China\\
$^{50}$ State Key Laboratory of Particle Detection and Electronics, Beijing 100049, Hefei 230026, People's Republic of China\\
% ...
$^{58}$ University of Jinan, Jinan 250022, People's Republic of China\\
% ...
$^{64}$ University of Science and Technology of China, Hefei 230026, People's Republic of China\\
% ...
$^{67}$ University of Turin and INFN, (A)University of Turin, I-10125, Turin, Italy; (B)University of Eastern Piedmont, I-15121, Alessandria, Italy; (C)INFN, I-10125, Turin, Italy\\
$^{68}$ Uppsala University, Box 516, SE-75120 Uppsala, Sweden\\
% ...
$^{b}$ Also at the Novosibirsk State University, Novosibirsk, 630090, Russia\\
% ...
}\vspace{0.4cm}}} 


\begin{abstract}
We present a measurement of the dressed cross sections for $e^+ e^- \rightarrow \phi \eta^{\prime}$ at different center-of-mass energies between 3.508 and 4.600 GeV based on 15.1 fb$^{-1}$ of $e^+ e^-$ annihilation data collected with the BESIII detector operating at the BEPCII collider. In addition, a search for the decay $Y(4230) \to \phi \eta^{\prime}$ is performed. No clear signal is observed and the corresponding upper limit is provided. 
\end{abstract}

\maketitle


\section{Introduction}
With the observation of a series of new charmoniumlike states, the so-called $XYZ$ states, charmonium physics has seen a resurgence of interest from both theory and experiment. These charmoniumlike states do not fit in the conventional $c\bar{c}$ charmonium spectroscopy and could be exotic states that lie outside the naive quark model~\cite{Brambilla:2010cs}. A better understanding of these states would shed light on the nonperturbative regime of the strong interaction. The first observed $Y$ state, the $Y(4260)$, was found by  the $BaBar$ collaboration in the initial state radiation (ISR) process $e^+e^-\to\gamma_\text{ISR} J/\psi\pi^+\pi^-$~\cite{Aubert:2005rm}; it was confirmed by CLEO-c~\cite{He:2006kg}, Belle~\cite{Yuan:2007sj} and in an updated analysis by the $BaBar$ collaboration~\cite{Lees:2012cn}. In later experiments, the $Y(4260)$ was also observed in a series of processes measured by the BESIII collaboration, such as $e^+ e^- \to \pi^0\pi^0J/\psi$ ~\cite{Coan:2006rv}, $\omega\chi_{c0}$~\cite{Ablikim:2015uix}, $\pi^+\pi^-h_c$~\cite{BESIII:2016adj}, $\pi^+\pi^- J/\psi$~\cite{Ablikim:2016qzw}, $\pi^+\pi^-\psi(3686)$~\cite{Ablikim:2017oaf}, and $\pi^+D^0D^{*-}$~\cite{Ablikim:2018vxx}. Furthermore, evidence for transitions from the $Y(4260)$ to other charmoniumlike states, such as the $X(3872)$ and $Z_c(3900)$, have been reported~\cite{Ablikim:2019zio, BESIII:2020pov}. These new measurements at the BESIII experiment also led to a downward shift in the mass of the $Y(4260)$, so it has been renamed the $Y(4230)$~\cite{ParticleDataGroup:2022pth}. In the remainder of this paper, we will use $Y(4230)$ to represent this state\footnote{The Particle Data Group also calls this state the $\psi(4230)$, according to its quantum numbers $J^{PC} = 1^{--}$}. 

The internal structure of the $Y(4230)$ state remains controversial and many theoretical models have been proposed to interpret the $Y(4230)$, as a heavy charmonium state~\cite{Dubynskiy:2008mq,Li:2013ssa}, a tetraquark state~\cite{Ebert:2008kb}, a hadronic molecule~\cite{Cleven:2013mka,Ding:2008gr,Li:2013yla,Wang:2013cya,Wang:2013onz} or hybrid charmonium~\cite{Liu:2012ze, Chen:2016ejo}, but none of them has been conclusive. Searches for new decay modes of the $Y(4230)$ will provide more information that can help us understand its production and decay mechanisms, and reveal its structure. In addition to the processes mentioned before, several analyses have been performed to search for the decays of the $Y(4230)$ into light hadron final states, such as $e^+ e^- \to$ $K^{+} K^{-} \pi^0/\eta$~\cite{Aubert:2007ym}, $\phi \pi^+ \pi^-$\cite{Aubert:2006bu}, $\phi \phi \phi$, $\phi \phi \omega$~\cite{Ablikim:2017rnw}, $p \bar{p} \pi^0$~\cite{Ablikim:2017gtb}, $\eta Y(2175)$~\cite{Ablikim:2017auj}, $K_{S}^{0}K^{\pm}\pi^{\mp}\pi^0/\eta$~\cite{Ablikim:2018ddb}, $K_{S}^{0}K^{\pm}\pi^{\mp}$~\cite{Ablikim:2018jbb} and so on, but none of these measurements has observed obvious $Y(4230)$ signal. 

In this paper, we utilize data samples collected by the BESIII detector to search for a new $Y(4230)$ decay mode by measuring the dressed cross sections for $e^{+} e^{-} \to \phi \eta^{\prime}$ at center-of-mass energies between 3.508 and 4.600~GeV, as summarized in Table~\ref{tab:cs}. This measurement is also an extension of a previous measurement of the same process in a lower energy region, performed in the vicinity of the $\phi(2170)$ by the BESIII collaboration~\cite{Ablikim:2020coo}. The $\phi$ meson is reconstructed through its $K^+K^-$ decay mode, and the $\eta^{\prime}$ through both its $\gamma \pi^+ \pi^-$ decay (mode I) and its decay to $\eta \pi^+ \pi^-$ with $\eta \to \gamma \gamma$ (mode II). The sum of data or Monte Carlo (MC) simulated samples at all 20 energy points are used hereafter unless explicitly stated.


\section{BESIII DETECTOR AND DATA SAMPLES}

The BESIII detector is a magnetic spectrometer~\cite{Ablikim:2009aa} located at the Beijing Electron Positron Collider (BEPCII)~\cite{Yu:IPAC2016-TUYA01}. The
cylindrical core of the BESIII detector consists of a helium-based multilayer drift chamber (MDC), a plastic scintillator time-of-flight system (TOF), and a CsI(Tl) electromagnetic calorimeter (EMC), which are all enclosed in a superconducting solenoidal magnet providing a 1.0~T magnetic field. The solenoid is supported by an octagonal flux-return yoke with resistive plate counter muon identifier modules interleaved with steel. The acceptance of charged particles and photons is 93\% over $4\pi$ solid angle. The charged-particle momentum resolution at $1~{\rm GeV}/c$ is ~$0.5\%$, and the resolution of specific ionization measured in the MDC ($dE/dx$) is $6\%$ for the electrons from Bhabha scattering. The EMC measures photon energies with a resolution of $2.5\%$ ($5\%$) at $1$~GeV in the barrel (end cap) region. The time resolution of the TOF barrel part is 68~ps, while that of the end cap part is 110~ps. The end cap TOF system was upgraded in 2015 with multigap resistive plate chamber technology, providing a time resolution of 60~ps~\cite{etof}. 

MC simulated data samples are used to determine the detection efficiency and to estimate the backgrounds. They are produced with a {\sc geant4}-based~\cite{geant4} software package that includes the geometric description of the BESIII detector and the detector response. To study the efficiency of each final state, a sample of $1 \times 10^5$ events is generated at each energy point, and the simulation includes the beam energy spread and ISR in the $e^+e^-$ annihilations modeled with {\sc kkmc}~\cite{ref:kkmc}. For the signal process, $e^{+} e^{-} \to \phi \eta^{\prime}$ is generated using the {\sc vvs\_pwave} decay model~\cite{Ping:2008zz, Lange:2001uf}; $\eta^{\prime} \to \gamma \pi^+ \pi^-$ and $\phi \to K^+ K^-$ are simulated with the $\rho^0-\omega-\mbox{box~anomaly}$ model considered~\cite{BESIII:2017kyd} and the {\sc vss} decay model~\cite{Ping:2008zz, Lange:2001uf}, respectively; other decay modes are generated with phase space distributions. For the background study, three generic MC samples at the energies of 3.773, 4.178 and ~4.226 GeV are generated. The known decay modes of the charmonium states are produced with {\sc evtgen}~\cite{Ping:2008zz, Lange:2001uf} according to the world average branching fraction (BF) values~\cite{ParticleDataGroup:2022pth}, while the unknown decay modes are generated with {\sc lundcharm}~\cite{Chen:2000tv,Yang:2014vra}. Final state radiation from charged final state particles is incorporated using the {\sc photos} package~\cite{Richter-Was:1992hxq}. Continuum hadronic events are generated with {\sc kkmc}~\cite{ref:kkmc} and QED processes such as Bhabha scattering, $\mu^+\mu^-$, $\tau^+\tau^-$, and $\gamma\gamma$ events are generated with {\sc kkmc}~\cite{ref:kkmc} and {\sc babayaga}~\cite{Balossini:2006wc}.


\section{Data Analysis}
\subsection{Event selections and background analysis}    
Charged tracks are required to have a polar angle $\theta$ with respect to the detector axis within the MDC acceptance $|\cos\theta| < 0.93$, and a distance of closest approach to the interaction point within 10~cm along the beam direction and 1~cm in the plane perpendicular to the beam direction. The particle type for each charged track is determined by selecting the hypothesis with the highest probability, which is calculated with the combination of time information from the TOF and $dE/dx$ for different particle hypotheses.

Photon candidates are reconstructed from isolated electromagnetic showers in the EMC. The energy of a photon candidate is required to be larger than 25 MeV (50 MeV) in the barrel (end cap) region, corresponding to an angular coverage of $|\cos\theta| < 0.80$ ($0.86 < |\cos\theta| < 0.92$). The electromagnetic shower time from the EMC has to be within 700 ns of the event start time to suppress electronic noise and energy deposition unrelated to the event of interest. To eliminate the showers associated with charged particles, the opening angle between a photon candidate and the extrapolated position of the closest charged track must be larger than 10 degrees.

The signal candidates for the $e^+ e^- \to \phi \eta^{\prime}$ process are selected by requiring four charged tracks with net charge zero and identified as $K^+$, $K^-$, $\pi^+$ and $\pi^-$, as well as at least one (two) photon(s) for mode I (II). To improve the resolution and suppress backgrounds, a four-constraint kinematic fit is performed for the decay mode I, constraining the total four-momentum of the final-state particles to the total initial four-momentum of the colliding beams. 
For the mode II, a five-constraint kinematic fit is performed with an additional constraint of the invariant mass of the two photons to the world average $\eta$ mass~\cite{ParticleDataGroup:2022pth}. If there is more than one combination in an event, the one with the smallest kinematic fit $\chi^2$ is selected. The $\chi^2$ of the candidate events are required to be less than 50 for both modes.

\begin{figure}[htbp]
\subfigure{
	\includegraphics[width = 0.39\textwidth]{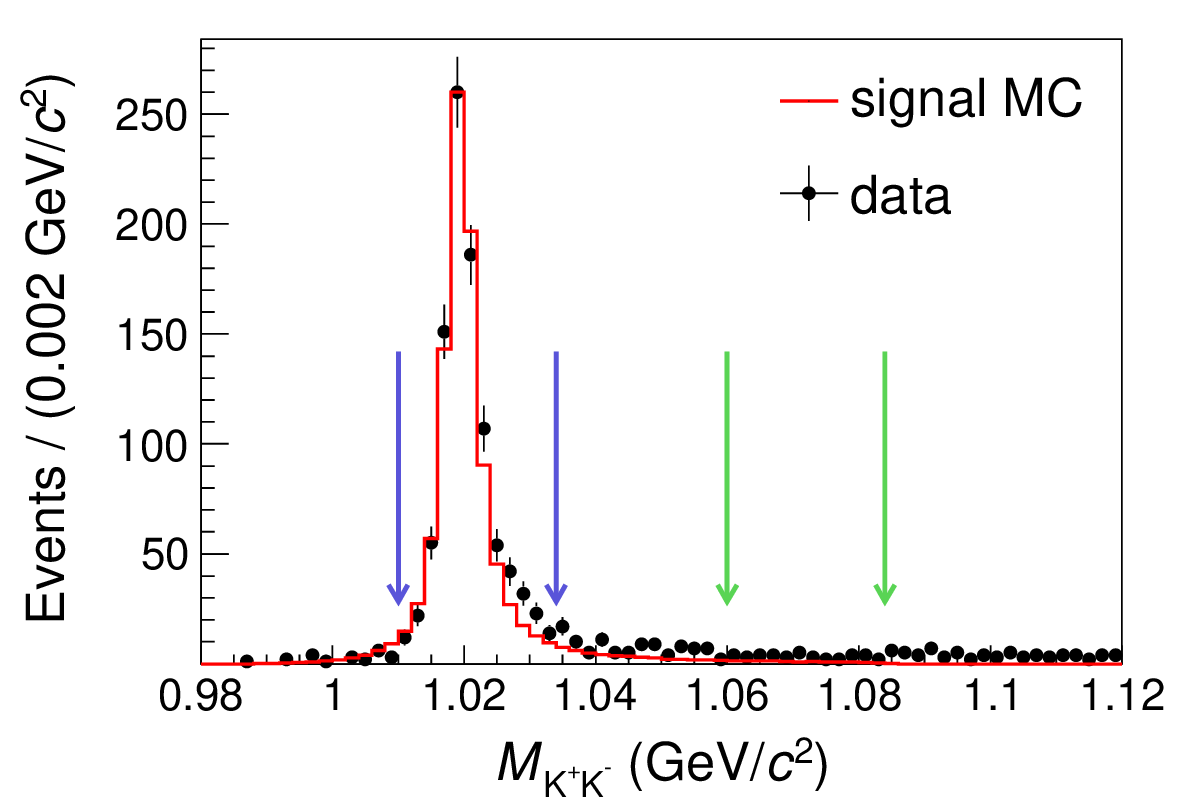}\put(-150,100){\footnotesize (a) }
}
\subfigure{
	\includegraphics[width = 0.39\textwidth]{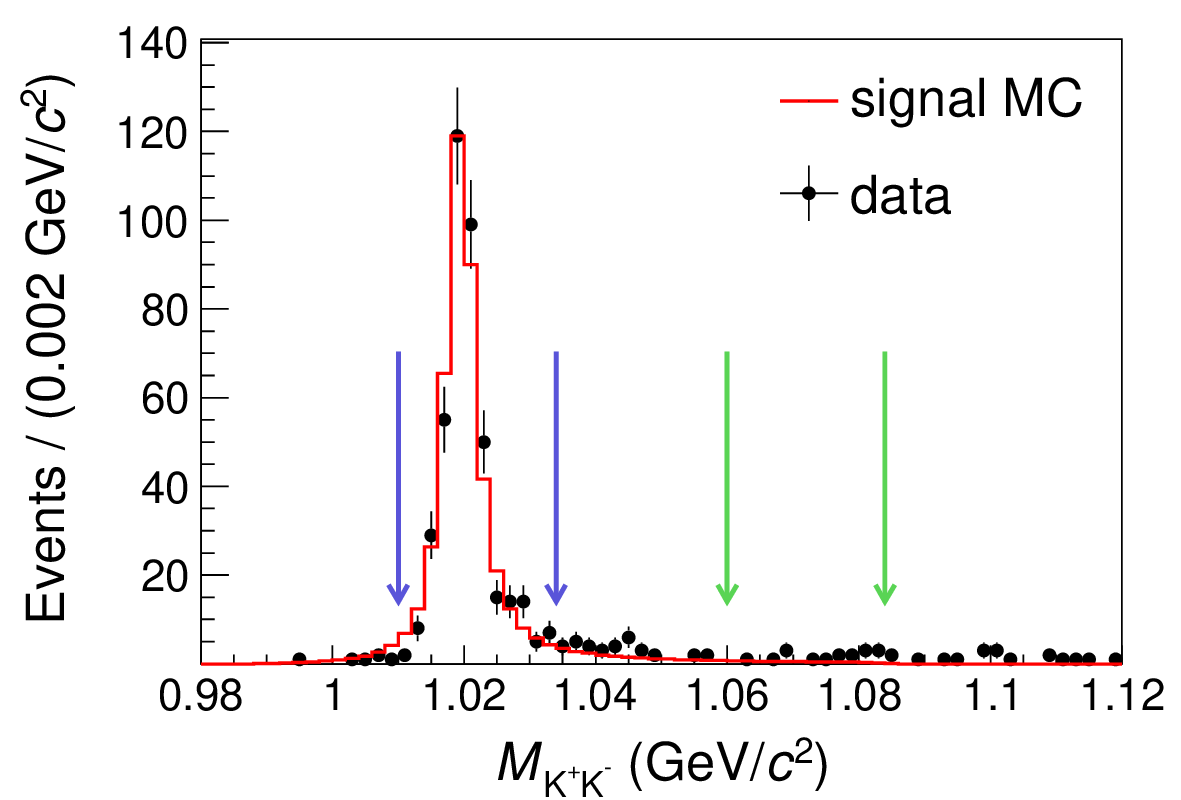}\put(-150,100){\footnotesize (b) }
}
\subfigure{
	\includegraphics[width = 0.39\textwidth]{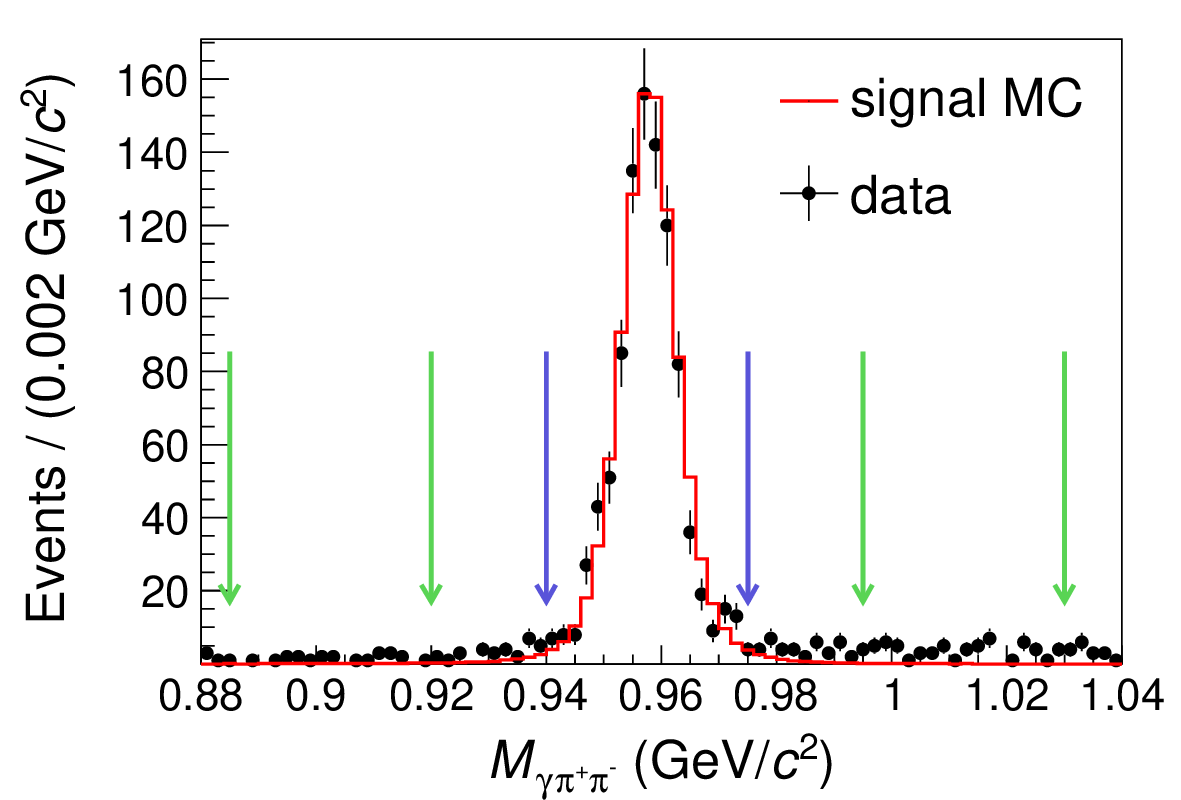}\put(-150,100){\footnotesize (c) }
}
\subfigure{              
\includegraphics[width = 0.39\textwidth]{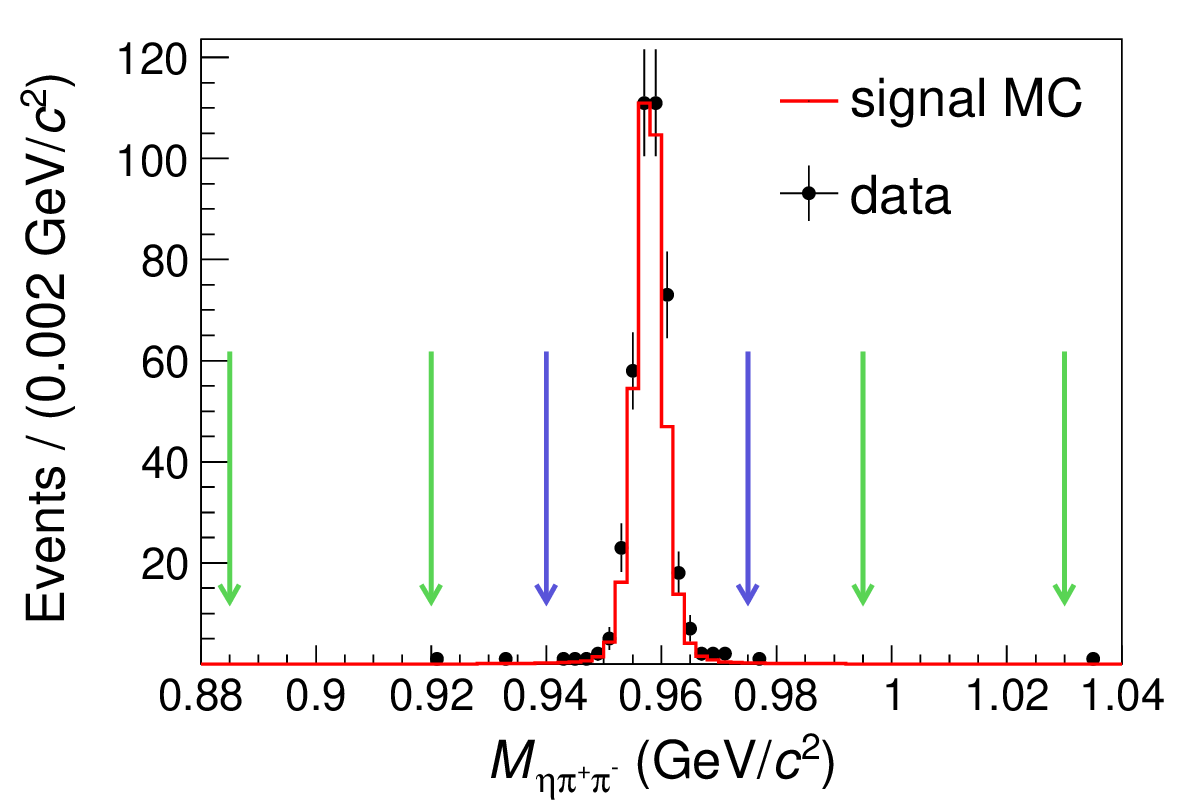}\put(-150,100){\footnotesize (d) }
}
\caption{Invariant mass spectra of the $\phi$ candidates after selecting the $\eta^{\prime}$ signal regions for mode I (a) and for mode II (b), and the invariant mass spectra for the $\eta^{\prime}$ candidates after selecting the $\phi$ signal regions for mode I (c) and for mode II (d). The black dots with error bars are experimental data, the red line is the signal MC, the slate blue arrows mark the signal regions, and the lime green arrows denote the sidebands. The MC simulation is arbitrarily normalized to data.}
\label{fig1}
\end{figure}

The signal candidates for the $\eta$, $\phi$, and $\eta^{\prime}$ mesons are selected within the mass ranges (in GeV/c$^2$) $0.500 \le M_{\gamma\gamma} \le 0.570$, $1.010 \le M_{K^+K^+} \le 1.034$ for both modes and $0.940 \le M_{\gamma \pi^+ \pi^-}(M_{\eta \pi^+ \pi^-}) \le 0.975$ for mode I (II). The sideband regions (also in GeV/$c^2$), defined as $M_{K^+K^-} \in [1.060, 1.084]$ for both modes and $M_{\gamma \pi^+ \pi^-} (M_{\eta \pi^+ \pi^-}) \in [0.885, 0.920] \cup [0.995, 1.030]$ for mode I (II) are used to estimate the backgrounds in signal regions, as shown in Fig.~\ref{fig1}.

\begin{figure}[htbp]
\subfigure{
	\includegraphics[width = 0.4\textwidth]{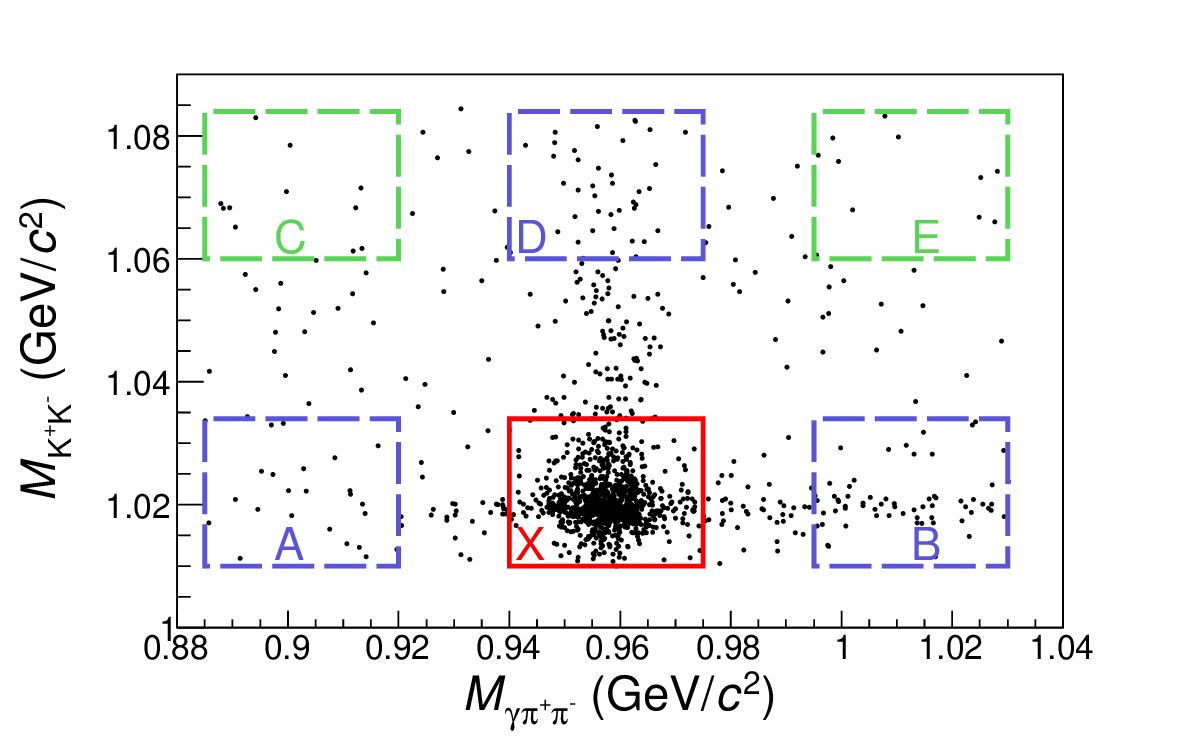}\put(-33,65){\footnotesize (a)}
}
\subfigure{
	\includegraphics[width = 0.4\textwidth]{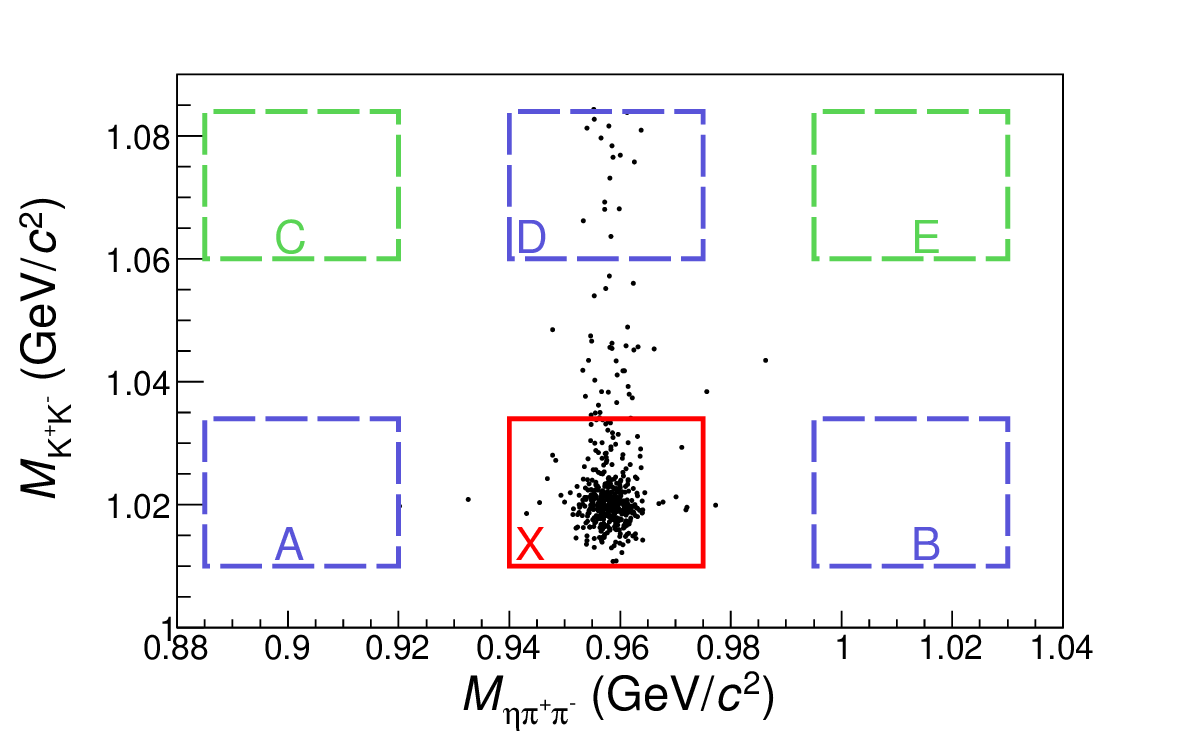}\put(-33,65){\footnotesize (b)}
}

\caption{Distributions of $M_{K^+K^-}$ versus $M_{\gamma \pi^+ \pi^-}$ for mode I (a) and $M_{K^+K^-}$ versus $M_{\eta \pi^+ \pi^-}$ for mode II (b), where the red rectangle shows the signal region, and the slate blue and lime green rectangles show the 2D sidebands.}
\label{fig2}
\end{figure}

The signal yields $N$ are determined from the distributions of $M_{K^+K^-}$ versus $M_{\gamma \pi^+ \pi^-}$ for mode I (a) and $M_{K^+K^-}$ versus $M_{\eta \pi^+ \pi^-}$ for mode II (b) by
\begin{equation}
N = N_{\rm X} - (N_{\rm A} + N_{\rm B})/2 - r\cdot N_{\rm D} + r\cdot (N_{\rm C} + N_{\rm E})/2, 
\label{eq:Nsig}
\end{equation}
where $N_{\rm X}$, $N_{\rm A}$, $N_{\rm B}$, $N_{\rm C}$, $N_{\rm D}$, and $N_{\rm E}$ represent the numbers of events observed in the equal-area regions X, A, B, C, D, and E, as shown in Fig.~\ref{fig2}. The boundaries of these regions correspond to the previous definitions of the signal and sidebands. The backgrounds due to misreconstruction of the $\eta^{\prime}$ are assumed to be linear in the $M_{\gamma \pi^+ \pi^-}$ and $M_{\eta \pi^+ \pi^-}$ distributions, and are estimated using the number of events in the regions A and B. The background with correctly reconstructed $\eta^{\prime}$ but no $\phi$ is estimated using the $M_{K^+K^-}$ sideband region D. The regions C and E represent the nonresonant background without a $\phi$ or an $\eta^{\prime}$. 

\begin{figure}[htbp]
\subfigure{
	\includegraphics[width = 0.35\textwidth]{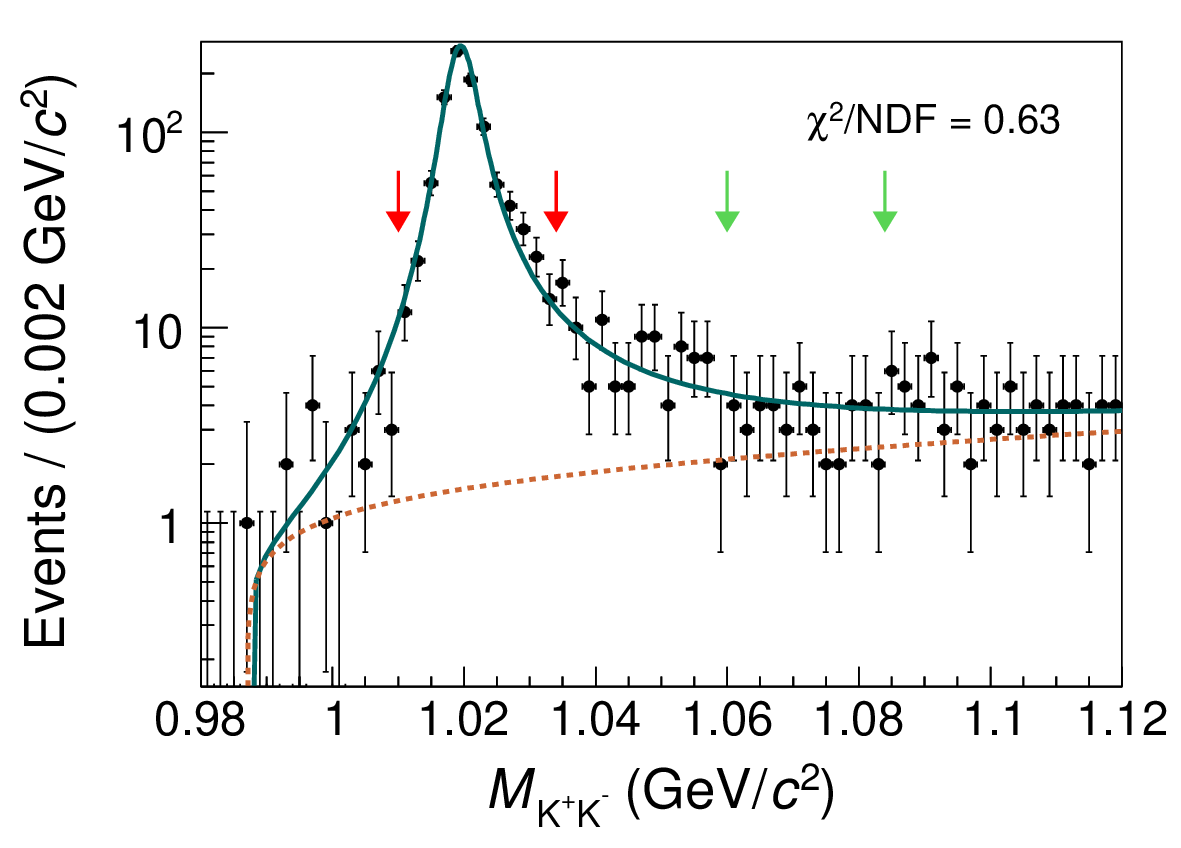}\put(-135,110){\footnotesize (a)}
}
\subfigure{
	\includegraphics[width = 0.35\textwidth]{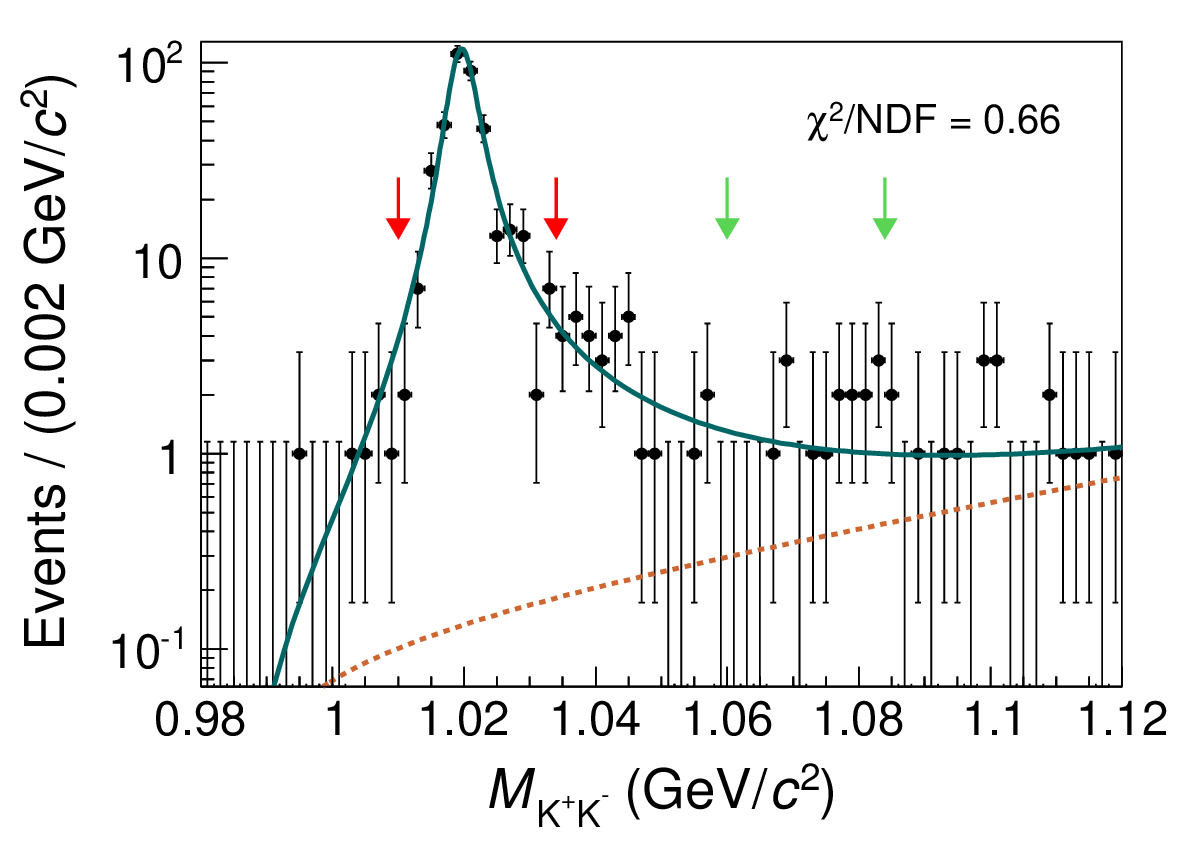}\put(-135,110){\footnotesize (b)}
}
\caption{Fits to the $M_{K^+K^-}$ distributions of the candidate events for mode I (a) and mode II (b). The black dots with error bars are data. The solid curves denote the total fits, where the dotted lines represent the background, the two red arrows and two lime green arrows show the signal and sideband regions, respectively.}
\label{fig3}
\end{figure}

The ratio of non-$\phi$ backgrounds under the $M_{K^+K^-}$ peak over that in the sideband region of $M_{K^+K^-}$ is defined as $r$, which is evaluated to be 0.66 and 0.39 for mode I and mode II, respectively. To obtain the value of $r$ for the two modes, maximum likelihood fits are performed on the $M_{K^+K^-}$ distributions for the two modes, as shown in Fig.~\ref{fig3}. The probability density function of the $M_{K^+K^-}$ spectra for the $\phi$ is obtained from a $P$-wave Breit-Wigner function convolved with a Gaussian function that accounts for the detector resolution. The $P$-wave Breit-Wigner function is defined as

\begin{equation}
\mathit {f}(m)=|\mathrm{A}(m)|^{2}\cdot p,
\end{equation}

\begin{equation}
\mathrm{A}(m)=\frac{p^{\ell}}{m^2-m^{2}_{0}+im\Gamma(m) }
\cdot\frac{\mathrm{B}(p)}{\mathrm{B}(p^{'})},
\end{equation}

\begin{equation}
\mathrm{B}(p)=\frac{1}{\sqrt{1+(\mathrm{R}p)^{2}}},
\end{equation}

\begin{equation}
\Gamma(m)=\left(\frac{p}{p^{'}}\right)^{2\ell + 1}\left(\frac{m_{0}}{m}\right)\Gamma_{0}\left[\frac{\mathrm{B}(p)}{\mathrm{B}(p^{'})}\right],
\end{equation}

\noindent where $m_{0}$ is the nominal $\phi$ mass as specified in the Particle Data Group~\cite{ParticleDataGroup:2022pth}, 
$p$ is the momentum of the kaon in the rest frame of the $K^+K^-$
system, $p^{'}$ is the momentum of the kaon at the nominal mass of the
$\phi$, and $\Gamma_{0}$ is the width of the $\phi$. The angular
momentum ($\ell$) is assumed to be equal to one, which is the lowest allowed
given the parent and daughter spins, $\mathrm{B}(p)$ is the
Blatt-Weisskopf form factor, and $R$ is the radius of the centrifugal
barrier, whose value is taken to be 3~{GeV/\it{c}}$^{-1}$~\cite{Blatt-Wdisskopf}.

The background shape is described by an ARGUS
function~\cite{Albrecht:1990am}. The parameters of the Gaussian function
and the ARGUS function are left free in the fit. The data is described by an incoherent sum of signal and background contributions.

To validate our analysis method, we also perform the same treatment for generic MC samples to validate the reliability of the method at the energies 3.773 and 4.178 GeV. The signal obtained from Eq.~(\ref{eq:Nsig}) is consistent with zero, as expected for the background-only generic MC sample.

\begin{table*}
\begin{center}
	\caption{Summary of the integrated luminosities $\mathcal{L}$, the detection efficiencies $\epsilon$,  the numbers of events in the signal region $N^{\rm sig}$,  the ISR correction factors $1+\delta^{\rm ISR}$, the measured dressed cross sections $\sigma^{\rm dre}$ for individual modes, and the combined dressed cross sections $\sigma^{\rm dre}_{\rm  comb}$, where subscripts I and II represent modes I and II, respectively. Only statistical uncertainties are shown.}
	\label{tab:cs}
	\scalebox{1.0}{
	\begin{tabular}{lllllllllll}
		\hline\hline
		$\sqrt{s}~(\rm GeV)$   &$\mathcal{L}~(\rm pb^{-1})$      &$\epsilon_{\rm I}~(\%)$  &$\epsilon_{\rm II}~(\%)$   &$N_{\rm I}^{\rm sig}$   &$N_{\rm II}^{\rm sig}$   &$1+\delta^{\rm ISR}$    &$\sigma^{\rm dre}_{\rm I}~(\rm pb)$  &$\sigma^{\rm dre}_{\rm II}~(\rm pb)$     &$\sigma^{\rm dre}_{\rm  comb}~(\rm pb)$  \\
		\hline
		3.50800   &181.8   &$16.93$   &$15.74$     &$22.2^{+ 5.8}_{- 4.8}$         &$11.0^{+ 4.2}_{- 3.0}$           &$1.022$         &$4.96^{+ 1.30}_{- 1.07}$   &$4.57^{+ 1.75}_{- 1.25}$        &$4.82^{+ 1.01}_{- 0.85}$\\
		3.51060    &184.6  &$16.98$   &$15.59$     &$34.0^{+ 6.2}_{- 6.1}$         &$12.1^{+ 4.3}_{- 3.2}$           &$1.022$         &$7.47^{+ 1.35}_{- 1.34}$   &$4.98^{+ 1.78}_{- 1.30}$        &$6.63^{+ 0.51}_{- 0.13}$\\
		3.77300   &2931.8 &$17.83$   &$16.67$     &$295.1^{+ 18.5}_{- 18.4}$      &$124.5^{+ 11.5}_{- 11.1}$        &$1.079$         &$3.68^{+ 0.23}_{- 0.23}$   &$2.86^{+ 0.26}_{- 0.25}$        &$3.34^{+ 0.18}_{- 0.18}$\\
		3.86741   &108.9    &$18.04$   &$17.01$     &$4.5^{+ 3.5}_{- 2.8}$          &$2.4^{+ 2.4}_{- 1.6}$            &$1.110$         &$1.44^{+ 1.13}_{- 0.91}$   &$1.41^{+ 1.44}_{- 0.94}$        &$1.43^{+ 0.85}_{- 0.69}$\\
		3.87131    &110.3    &$17.81$   &$16.63$     &$10.2^{+ 4.0}_{- 3.0}$         &$1.4^{+ 2.2}_{- 1.4}$            &$1.110$         &$3.30^{+ 1.29}_{- 0.96}$   &$0.83^{+ 1.28}_{- 0.82}$        &$2.34^{+ 0.23}_{- 0.23}$\\
		4.00762   &482.0   &$17.67$   &$16.80$     &$27.1^{+ 6.0}_{- 5.8}$         &$15.9^{+ 4.9}_{- 3.6}$           &$1.146$         &$1.95^{+ 0.43}_{- 0.42}$   &$2.07^{+ 0.64}_{- 0.46}$        &$2.00^{+ 0.34}_{- 0.31}$\\
		4.17800   &3194.5  &$17.29$   &$16.57$     &$158.0^{+ 13.2}_{- 13.1}$      &$61.0^{+ 8.0}_{- 8.0}$           &$1.192$         &$1.69^{+ 0.14}_{- 0.14}$   &$1.17^{+ 0.15}_{- 0.15}$        &$1.46^{+ 0.10}_{- 0.10}$\\
		4.18899   &523.9   &$17.23$   &$16.50$     &$29.1^{+ 6.1}_{- 6.0}$         &$6.0^{+ 3.5}_{- 2.2}$            &$1.198$         &$1.89^{+ 0.40}_{- 0.39}$   &$0.70^{+ 0.40}_{- 0.26}$        &$1.50^{+ 0.34}_{- 0.34}$\\
		4.19903   &525.2   &$17.66$   &$16.72$     &$22.1^{+ 5.7}_{- 4.6}$         &$9.1^{+ 3.9}_{- 2.8}$            &$1.199$         &$1.40^{+ 0.36}_{- 0.29}$   &$1.05^{+ 0.45}_{- 0.32}$        &$1.29^{+ 0.21}_{- 0.17}$\\
		4.20925   &517.2   &$16.88$   &$16.47$     &$27.0^{+ 6.2}_{- 5.0}$         &$6.5^{+ 3.2}_{- 2.5}$            &$1.203$         &$1.81^{+ 0.41}_{- 0.33}$   &$0.77^{+ 0.38}_{- 0.30}$        &$1.48^{+ 0.22}_{- 0.22}$\\
		4.21884   &513.4    &$16.73$   &$16.12$     &$25.1^{+ 5.9}_{- 4.8}$         &$10.1^{+ 4.0}_{- 2.9}$           &$1.208$         &$1.70^{+ 0.40}_{- 0.33}$   &$1.22^{+ 0.49}_{- 0.35}$        &$1.55^{+ 0.17}_{- 0.16}$\\
		4.22626   &1056.4  &$17.90$   &$16.74$     &$44.1^{+ 7.1}_{- 7.0}$         &$29.3^{+ 6.1}_{- 5.1}$           &$1.208$        &$1.36^{+ 0.22}_{- 0.22}$   &$1.66^{+ 0.34}_{- 0.29}$        &$1.47^{+ 0.11}_{- 0.11}$\\
		4.23582   &529.1    &$17.30$   &$16.52$     &$27.7^{+ 5.7}_{- 5.1}$         &$10.1^{+ 4.0}_{- 2.9}$           &$1.214$         &$1.75^{+ 0.36}_{- 0.32}$   &$1.15^{+ 0.46}_{- 0.33}$        &$1.56^{+ 0.14}_{- 0.14}$\\
		4.24393   &536.3    &$17.34$   &$16.60$     &$19.2^{+ 5.3}_{- 4.3}$         &$11.0^{+ 4.2}_{- 3.0}$           &$1.213$         &$1.20^{+ 0.33}_{- 0.27}$   &$1.24^{+ 0.47}_{- 0.34}$        &$1.21^{+ 0.26}_{- 0.22}$\\
		4.25797   &828.4    &$17.49$   &$16.72$     &$38.0^{+ 6.4}_{- 6.4}$         &$14.4^{+ 4.5}_{- 3.6}$           &$1.218$         &$1.51^{+ 0.26}_{- 0.26}$   &$1.03^{+ 0.32}_{- 0.26}$        &$1.34^{+ 0.08}_{- 0.08}$\\
		4.26680   &529.7    &$17.44$   &$16.69$     &$19.1^{+ 5.4}_{- 4.2}$         &$11.0^{+ 4.2}_{- 3.0}$           &$1.220$         &$1.19^{+ 0.33}_{- 0.26}$   &$1.24^{+ 0.47}_{- 0.34}$        &$1.21^{+ 0.26}_{- 0.22}$\\
		4.27774   &175.2     &$16.78$   &$16.26$     &$1.4^{+ 2.2}_{- 1.4}$          &$1.4^{+ 2.2}_{- 1.4}$            &$1.223$         &$0.27^{+ 0.42}_{- 0.27}$   &$0.48^{+ 0.75}_{- 0.48}$        &$0.34^{+ 0.36}_{- 0.26}$\\
		4.35826   &543.9    &$17.51$   &$16.73$      &$27.4^{+ 5.8}_{- 5.0}$        &$6.2^{+ 3.4}_{- 2.3}$            &$1.245$         &$1.62^{+ 0.34}_{- 0.29}$   &$0.66^{+ 0.36}_{- 0.25}$        &$1.33^{+ 0.20}_{- 0.20}$\\
		4.41558   &1043.9   &$17.83$   &$16.71$     &$42.1^{+ 7.1}_{- 7.0}$         &$11.0^{+ 4.2}_{- 3.0}$           &$1.268$         &$1.25^{+ 0.21}_{- 0.21}$   &$0.60^{+ 0.23}_{- 0.16}$        &$1.04^{+ 0.18}_{- 0.18}$\\
		4.59953   &586.9    &$17.54$   &$16.19$      &$14.2^{+ 4.8}_{- 3.7}$        &$7.8^{+ 3.9}_{- 2.5}$            &$1.334$         &$0.73^{+ 0.24}_{- 0.19}$   &$0.74^{+ 0.37}_{- 0.24}$        &$0.73^{+ 0.20}_{- 0.16}$\\
		\hline\hline
	\end{tabular}
	}
\end{center}
\end{table*}

To determine the statistical uncertainties of the number of signal events, we construct a likelihood function $L(s)$ as Eq.~(\ref{eq:CombLike}), where the number of events in the signal and sideband regions are $N_{\rm X}$ and $(N_{\rm X}-N)$, respectively, which both follow the Poisson distributions $P(N_{\rm X};x)$ and $P((N_{\rm X}-N);(x-s))$. Here $x$ and $(x-s)$ are the expected values of the number of events in the signal and sideband regions, respectively. The value of $s$ with the largest likelihood is taken as the estimated number of signal events $N^{\rm sig}$, and the asymmetric statistical uncertainties $\sigma^{+}_{\rm sig}$ and $\sigma^{-}_{\rm sig}$ can be obtained from the following formulas: 

\begin{equation}
L(s) = \int^{\infty}_{0} P(N_{\rm X};x)\cdot P((N_{\rm X}-N);(x-s))dx, 
\label{eq:CombLike}
\end{equation}

\begin{equation}
\int^{N^{\rm sig}}_{N^{\rm sig}-\sigma^-_{\rm  sig}}L(s)ds = 0.683\cdot \int^{N^{\rm sig}}_{-\infty}L(s)ds, 
\label{eq:sigmam}
\end{equation}  
\begin{equation}
\int^{N^{\rm sig}+\sigma^+_{\rm sig}}_{N^{\rm sig}}L(s)ds = 0.683\cdot \int^{\infty}_{N^{\rm sig}}L(s)ds. 
\label{eq:sigmap}
\end{equation}  

\subsection{Determination of $\sigma^{\rm dre}(e^+ e^- \rightarrow \phi \eta^{\prime})$}

The dressed cross section is calculated by
\begin{equation}\label{eq:sigma}
\sigma^{\rm{dre}} = \frac{N^{\mathrm{sig}}}{\mathcal{L}\cdot\mathcal{B}\cdot\epsilon\cdot(1+\delta^{\mathrm{ISR}})}\;,
\end{equation}
where $\mathcal{L}$ is the integrated luminosity; $\epsilon$ is the detection efficiency; $\mathcal{B}$ is the product of branching fractions, i.e. $\mathcal{B}(\phi \to K^+ K^-)\cdot \mathcal{B}(\eta^{\prime} \to \gamma \pi^+ \pi^-)$ for mode I, and $\mathcal{B}(\phi \to K^+ K^-) \cdot \mathcal{B}(\eta^{\prime} \to \eta \pi^+ \pi^-) \cdot \mathcal{B}(\eta \to \gamma \gamma)$ for mode II~\cite{ParticleDataGroup:2022pth}; $(1+\delta^{\rm{ISR}})$ is the ISR correction factor, which is obtained from a QED calculation taking the line shape of $e^{+}e^{-} \to \phi \eta^{\prime}$ cross sections at the 20 energy points shown in Table~\ref{tab:cs} as input, and is calculated in an iterative procedure until the variation of the correction factor is below 1\% compared to the previous iteration. The dressed cross sections are measured with asymmetric statistical uncertainties separately for the two decay modes $\eta^{\prime} \to \gamma \pi^+ \pi^-$ and $\eta^{\prime} \to \eta \pi^+ \pi^-$. The cross sections are taken as the parameters
of a variable Gaussian as defined in Ref.~\cite{Barlow:2004wg} to construct the likelihood function to give a combined result. The variable Gaussian form can be expressed as
\begin{equation}
\ln L(\sigma) = -\frac{1}{2}\sum_{i=1}^{2}\frac{(\sigma - \sigma_{i})^{2}}{\sigma^{+}_{i}\sigma^{-}_{i} + (\sigma^{+}_{i}-\sigma^{-}_{i})(\sigma - \sigma_{i})},  
\label{eq:CombXsLike}
\end{equation} 
where $\sigma$ represents the combined cross section, and $\sigma_{i}$, $\sigma_{i}^{+}$, and $\sigma_{i}^{-}$ are the cross sections and their asymmetric statistical uncertainties for the decay mode $i$. The asymmetric statistical uncertainties of combined cross section is obtained according to $\Delta \ln L = -1/2$. The results of the dressed cross section measurements are summarized in Table \ref{tab:cs}.


\subsection{Systematic uncertainties}

\begin{table}[htp]
\caption {\small Summary of relative systematic uncertainties (in percent) from the different sources.}
\centering 
\scalebox{0.85}{
	\begin{tabular}{l|c|c|c} \hline \hline
	Source   
	&\multicolumn{3}{c}{Relative uncertainty~($\%$)} \\ 
	\cline{2-4}
	& $\eta^{\prime} \to \gamma \pi^{+} \pi^{-}$ &$\eta^{\prime} \to \eta \pi^{+} \pi^{-}$  &Combination\\ \hline
	Center-of-mass energy                                &Negligible      &Negligible    &Negligible   \\
	Luminosity                                                   &1.0  &1.0  &1.0\\ 
	Tracking                                                     &4.0  &4.0  &4.0\\ 
	PID                                                          &4.0  &4.0  &4.0\\
	Photon                                                       &1.0  &2.0  &1.4\\
	Kinematic fit                                                &0.1  &0.8  &0.4\\ 
	ISR factor                                                   &1.0  &1.0  &1.0\\ 
	$\eta$ reconstruction                                        &$\cdots$    &1.0  &0.4\\
	Mass window of $\phi$                                        &1.7  &0.8  &1.4\\
	Mass window of $\eta^{\prime}$                               &1.1  &0.9  &1.0\\
	Sideband                                                     &0.3  &0.4  &0.3\\ 
	${\mathcal{B}}(\phi \to K^{+}K^{-})$                         &1.0  &1.0  &1.0\\ 
	${\mathcal{B}}(\eta^{\prime} \to \eta \pi^{+}\pi^{-})$       &$\cdots$    &1.6  &0.6\\  
	${\mathcal{B}}(\eta^{\prime} \to \gamma \pi^{+}\pi^{-})$     &1.7  &$\cdots$    &1.1\\
	${\mathcal{B}}(\eta \to \gamma \gamma)$                      &$\cdots$    &0.5  &0.2\\ 
	MC statistics                                                                 &0.7   &0.8  &0.5\\
	Total                                                        &6.6  &6.8  &6.5\\ \hline \hline
	\end{tabular}
}
\label{tab:combsys} 
\end{table}

The systematic uncertainties on the cross sections mainly come from the center-of-mass energy, luminosity, tracking efficiency, PID, photon and $\eta$ reconstruction, ISR correction factor, quoted BFs and kinematic fit, which are energy independent. The uncertainty of the center-of-mass energy is less than 0.01\%, measured by analyzing the dimuon process $e^+ e^- \to  \gamma_{\rm {ISR/FSR}}\mu^{+}\mu^{-}$, and is negligible~\cite{Ablikim:2015zaa}. The uncertainty on the integrated luminosity is estimated to be 1.0\% using events from large-angle Bhabha scattering~\cite{Ablikim:2015nan}. The tracking and PID differences in the efficiencies between data and MC simulation are studied using the control samples $J/\psi \to K^0_{S}K^{\pm}\pi^{\mp}$, $J/\psi \to \pi^+\pi^-\pi^0$, and $J/\psi \to K^+K^-\pi^0$, and the systematic uncertainties of tracking and PID are both determined to be 4\% (1\% per track)~\cite{Ablikim:2011kv}. The systematic uncertainty from the photon detection efficiency is determined to be 1\% per photon by utilizing a control sample of $J/\psi \to \rho^0\pi^0$ with $\rho^0 \to \pi^+\pi^-$ and $\pi^0 \to \gamma \gamma$~\cite{Ablikim:2010zn}. The uncertainty from the $\eta$ selection is 1\% per $\eta$, which is determined from a control sample of $J/\psi \to \eta p \bar{p}$~\cite{Ablikim:2010zn}. For the ISR correction factor, we use a power function to parametrize the line shape of the cross sections, then change the line shape by using or not using the low energy points from 2.900 to 3.080 GeV \cite{Ablikim:2020coo}, and take the difference in the cross sections as the systematic uncertainty. The uncertainties of the quoted BFs are taken from~\cite{ParticleDataGroup:2022pth}. The systematic uncertainty due to the kinematic fit is estimated by correcting the track helix parameters of charged tracks and the corresponding covariance matrix for the signal MC sample to improve the agreement between data and MC simulation. The detailed method can be found in Ref.~\cite{Ablikim:2012pg}. The resulting change of the detection efficiency with respect to the one obtained without the corrections is taken as the systematic uncertainty. 

The systematic uncertainties from the mass interval for the signal and sideband regions are estimated for all energy points with the largest data sample at 4.178 GeV. The systematic uncertainties associated with the signal regions of the $\phi$ and $\eta^{\prime}$ are estimated by changing the $\phi$ region from (1.010, 1.034) to (1.0076, 1.0364) and the $\eta^{\prime}$ region from (0.940, 0.975) to (0.93825, 0.97675) GeV/$c^2$. The uncertainties due to the sideband regions are determined by changing the mass windows to $M_{K^+K^-} \in (1.0588, 1.0852)$ GeV/$c^{2}$, $M_{\gamma\pi^+\pi^-}$ and $M_{\eta\pi^+\pi^-}\in(0.8832,0.9218)\cup(0.9933,1.0318)$ GeV/$c^2$. The differences in the efficiencies are taken as the corresponding systematic uncertainties. The uncertainty in MC statistics is obtained by $\frac{1}{\sqrt{N}}\sqrt{\frac{1-\epsilon}{\epsilon}}$, where the $\epsilon$ is detection efficiency and $N$ is the total number of the generated MC events.

\begin{figure}[htbp]
\subfigure{
	\includegraphics[width = 0.47\textwidth]{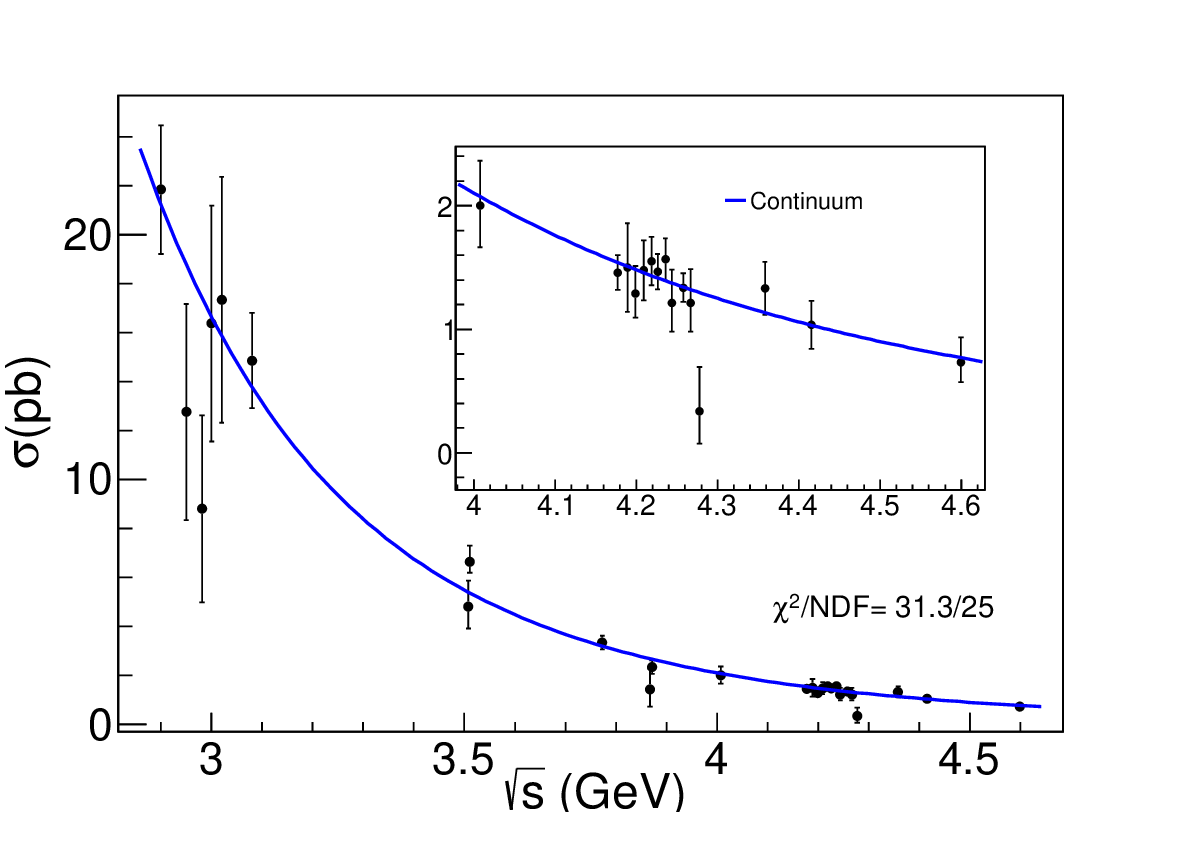}\put(-180,140){\footnotesize (a)}
}
\subfigure{
	\includegraphics[width = 0.47\textwidth]{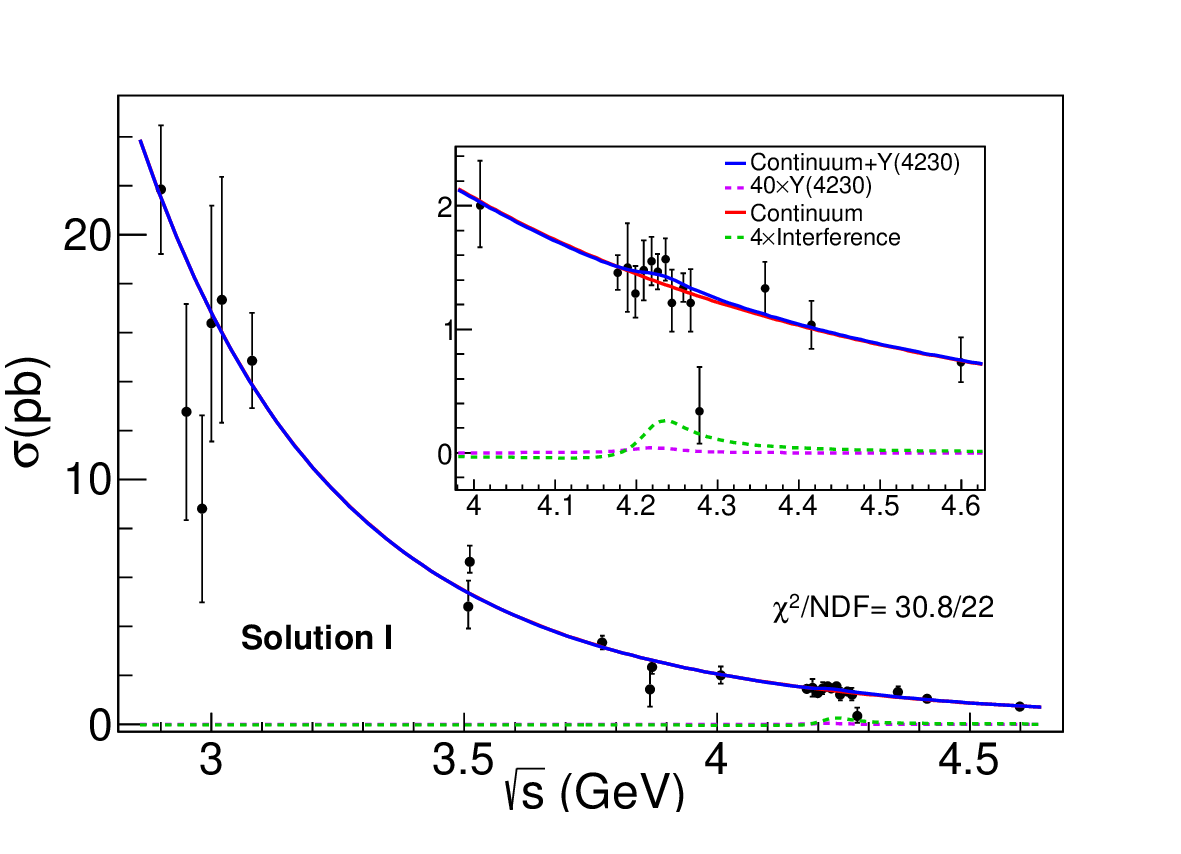}\put(-180,140){\footnotesize (b)}
}
\subfigure{
	\includegraphics[width = 0.47\textwidth]{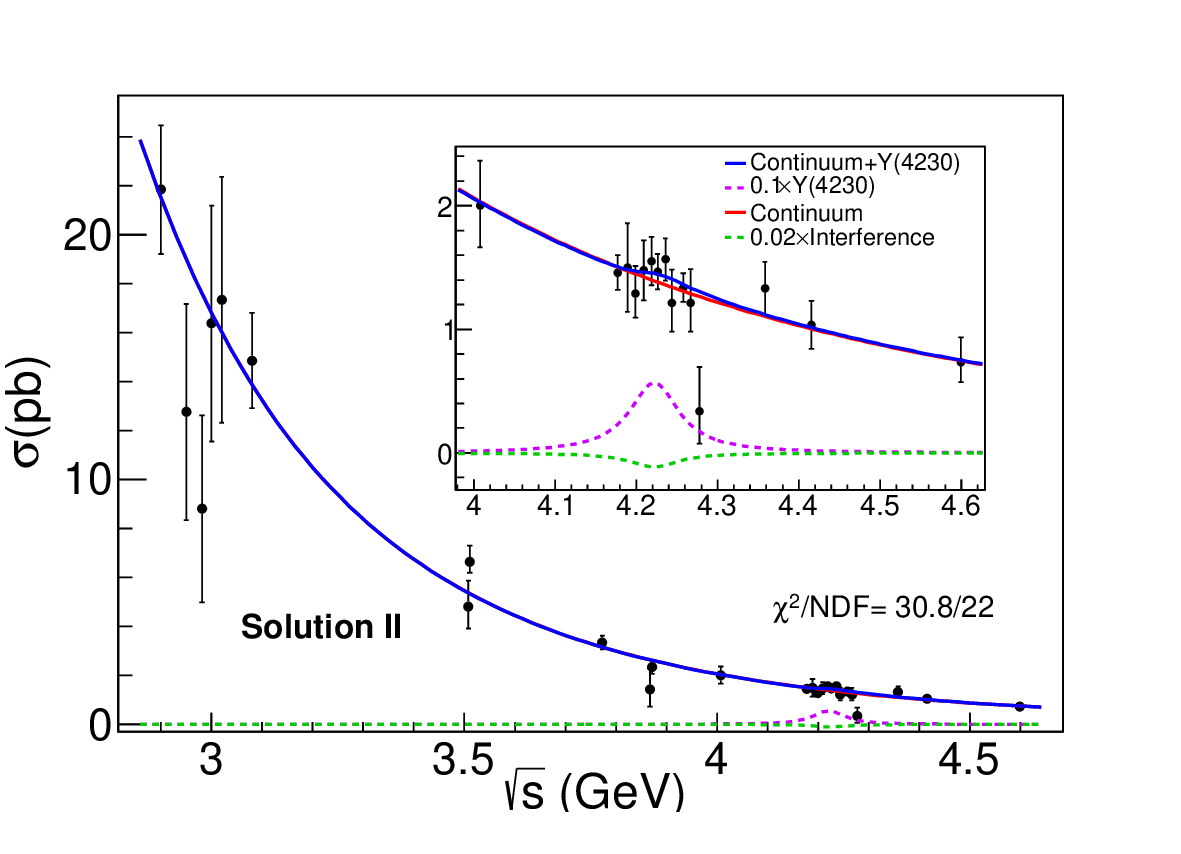}\put(-180,140){\footnotesize (c)}
}
\caption{Fit to the dressed cross sections of $e^+e^- \to \phi \eta^{\prime}$. The inset plots show the same distributions at center-of-mass energies between 4.009 and 4.600 GeV. The data (black dots) include both statistical and systematic uncertainties, the blue lines are the total fit results, the violet dashed lines are for the resonant components, the red solid lines are for the continuum process, the green dashed lines are for the interference parts, the violet and green dashed lines are scaled with an arbitrary factor to see them clearly. The plot (a) is the result with continuum process only, plots (b) and (c) are the results with continuum and $Y(4230)$ for solutions I and II, respectively.}
\label{lineshape}
\end{figure}

Due to correlations between the two decay modes, the combined systematic uncertainties $\zeta^{i}$ are calculated with 
\begin{equation} \label{eq:combsys}
\zeta = \frac{\sqrt{(w_{1}\zeta_{1})^{2}+(w_{2}\zeta_{2})^{2}+2w_{1}w_{2}\rho_{12}\zeta_{1}\zeta_{2}}}{w_{1}+w_{2}}, 
\end{equation}
where $w_{1}$, $w_{2}$ represent $\mathcal{B}(\eta^{\prime} \to \gamma \pi^{+} \pi^{-}) \cdot \epsilon_{\rm 1}$ and $\mathcal{B}(\eta^{\prime} \to \eta \pi^{+} \pi^{-}) \cdot \mathcal{B}(\eta \to \gamma \gamma) \cdot \epsilon_{\rm 2}$ respectively.  The corresponding uncertainties for modes I and II are denoted $\zeta_{1}$ and $\zeta_{2}$, respectively. The correlation coefficients $\rho_{12}$ are taken as 0 for items that are uncorrelated between the two modes, such as $\eta$ reconstruction, $\mathcal{B}(\eta^{\prime}\to \eta \pi^+ \pi^-)$, $\mathcal{B}(\eta^{\prime}\to \gamma \pi^+ \pi^-)$, $\mathcal{B}(\eta \to \gamma \gamma)$, and MC statistics. Other systematic effects are correlated between the two modes and $\rho_{12}$ is taken as 1. Table~\ref{tab:combsys} summarizes all the systematic uncertainties related to the cross section measurement of the $e^+e^- \to \phi \eta^{\prime}$ process for the individual decay modes and the combined one, respectively. The overall systematic uncertainties are obtained by adding all the sources of systematic uncertainties in quadrature, assuming they are all uncorrelated.

\section{Upper limit on $\Gamma_{ee}\times \mathcal{B}(Y(4230)\to \phi \eta^{\prime})$} 

Since there is no obvious structure in the line shape of the dressed cross sections of $e^{+}e^{-} \to \phi \eta^{\prime}$, as shown in Fig.~\ref{lineshape}, the upper limit of $Y(4230) \to \phi \eta^{\prime}$ is determined by fitting the line shape using a coherent sum of the continuum and the resonance $Y(4230)$ amplitude with Eq.~(\ref{eq:BW})
\begin{equation}\label{eq:BW}
\centering
\sigma^{\rm dre}(\sqrt{s}) = \Bigg |{ \sqrt{ \frac{f_{\rm con}}{s^{n}}} + e^{i\phi} \frac{\sqrt{12\pi \Gamma_{\rm ee} \mathcal{B}_{\phi \eta^{\prime}} \Gamma}}{s - M^2 + iM\Gamma} } \Bigg |^2, 
\end{equation}

where $f_{\rm con}$ and $n$ are the fit parameters for the continuum process; $\phi$ is the relative phase between the continuum and resonant amplitudes; $\Gamma$ and $\Gamma_{\rm ee}$ are the total width and partial width to $e^+ e^-$, respectively; $\mathcal{B}_{\phi \eta^{\prime}}$ is the branching fraction for the resonance decay into $\phi \eta^{\prime}$, and $M$ is the mass of the resonance. The mass and total width of the $Y(4230)$ are set to the world average values $4222.7 \pm 2.6$ and $49\pm8$~MeV~\cite{ParticleDataGroup:2022pth}. We vary the product $\Gamma_{\rm ee} \cdot\mathcal{B}_{\phi \eta^{\prime}}$ with fixed step size. For each value, the correlations among different data points are considered in the fit with a fitting estimator $Q^{2}$ constructed as Eq.~(\ref{eq:chi2}), which is minimized by {\sc minuit}~\cite{James:1975dr}.

\begin{equation}\label{eq:chi2}
\centering
Q^{2} = \sum_{i}\frac{(\sigma^{\rm dre_{i}}-h\cdot\sigma^{\rm dre_{i}}_{\rm fit})^{2}}{\delta_{i}^{2}} + \frac{(h-1)^{2}}{\delta_{c}^{2}}
\end{equation}       
Here $\sigma^{\rm dre_{i}}$ and $\sigma^{\rm dre_{i}}_{\rm fit}$ are the measured and fitted dressed cross section of the $i{\rm th}$ energy point, respectively; $\delta_{i}$ is the energy-dependent part of the total uncertainty at each energy point, which includes the statistical uncertainty and the energy-dependent contribution to the systematic uncertainty; $\delta_{c}$ is the relative systematic uncertainty corresponding to the energy-independent part, and $h$ is a free parameter introduced to take into account the correlations~\cite{Mo:2007aea}. Then we construct the likelihood by $L=e^{-0.5Q^{2}}$ , whose normalized distribution is used to get the upper limit of $\Gamma_{\rm ee}\cdot\mathcal{B}_{\phi \eta^{\prime}}$ at the 90\% C.L. Two solutions with the same minimum value of $Q^2$ are found with different interference between the two amplitudes, where the second solution can also be derived from the first solution by a numerical method~\cite{Zhu:2011ha}. The fit results are shown in Fig.~\ref{lineshape} (the line shapes of the two solutions are identical) and summarized in Table~\ref{tab:Solu}. Since we cannot determine which solutions are real, we only set the larger one as the upper limit of $\Gamma_{\rm ee} \cdot\mathcal{B}_{\phi \eta^{\prime}}$ to be 0.53 eV conservatively.

\begin{table*}[htpb]
\caption{Summary of the fit results to the measured dressed cross sections of $e^+ e^- \to \phi \eta^{\prime}$ with two solutions. The uncertainties of the parameters are from the fits.}
\begin{center}
	\begin{tabular}{cccccc} \hline \hline
	\multirow{2}{*}{}   &\multicolumn{2}{c}{$Y(4230)$ (Best Fit)} &\multicolumn{1}{c}{}  &\multicolumn{2}{c}{$Y(4230)$ (Upper Limit)}\\
	\cline{2-3} \cline{5-6}  
	&Solution I &Solution II   &$\qquad$  &Solution I &Solution II\\ \hline
	$\Gamma_{\rm ee} \mathcal{B}_{\phi \eta^{\prime}}~(\rm eV)$  & $(9.2 \pm 24.2) \times 10^{-5}$   &$0.51 \pm 0.03$   &   &$8.9  \times 10^{-4}$  &0.53\\ 
	$\phi~(\rm rad)$  &$0.86\pm1.08$   &$-1.59\pm0.02$  &    &-   &-\\   \hline \hline 
	\end{tabular}   
\end{center}
\label{tab:Solu}
\end{table*}


\section{Summary}
The dressed cross sections of $e^+ e^- \rightarrow \phi \eta^{\prime}$ are measured with data samples collected with the BESIII detector operating with the BEPCII collider at 20 energy points between $\sqrt{s} =$ 3.508 and 4.600 GeV. The line shape of the dressed cross sections is consistent with the continuum process. A fit with an additional resonance is performed to search for the decay $Y(4230)\to \phi \eta^{\prime}$. No clear resonant structure is observed, and an upper limit on the product $\Gamma_{ee}\times \mathcal{B}(Y(4230)\to \phi \eta^{\prime})$ at the 90\% C.L.\ is determined to be less than 0.53 eV.


\section*{Acknowledgements}
The BESIII collaboration thanks the staff of BEPCII and the IHEP computing center for their strong support. This work is supported in part by National Key R\&D Program of China under Contracts No. 2020YFA0406300, No. 2020YFA0406400; National Natural Science Foundation of China (NSFC) under Contracts No. 11625523, No. 11635010, No. 11735014, No. 11822506, No. 11835012, No. 11935015, No. 11935016, No. 11935018, No. 11961141012, No. 12022510, No. 12025502, No. 12035009, No. 12035013, and No. 12061131003; the Chinese Academy of Sciences (CAS) Large-Scale Scientific Facility Program; Joint Large-Scale Scientific Facility Funds of the NSFC and CAS under Contracts No. U1732263 and No. U1832207; CAS Key Research Program of Frontier Sciences under Contract No. QYZDJ-SSW-SLH040; 100 Talents Program of CAS; INPAC and Shanghai Key Laboratory for Particle Physics and Cosmology; ERC under Contract No. 758462; European Union Horizon 2020 research and innovation programme under Contract No. Marie Sklodowska-Curie Grant Agreement No. 894790; German Research Foundation DFG under Contracts No. 443159800, Collaborative Research Center CRC 1044, FOR 2359, GRK 2149; Istituto Nazionale di Fisica Nucleare, Italy; Ministry of Development of Turkey under Contract No. DPT2006K-120470; National Science and Technology fund; Olle Engkvist Foundation under Contract No. 200-0605; STFC (United Kingdom); The Knut and Alice Wallenberg Foundation (Sweden) under Contract No. 2016.0157; The Royal Society, UK under Contracts Nos. DH140054, DH160214; The Swedish Research Council; U. S. Department of Energy under Contracts Nos. DE-FG02-05ER41374, DE-SC-0012069.

\end{document}